\documentclass[conference]{IEEEtran}
\IEEEoverridecommandlockouts
\usepackage[colorinlistoftodos,prependcaption,textsize=tiny]{todonotes}
\usepackage{cite}
\usepackage{pgfplots}
\usepackage{amsmath,amssymb,amsfonts}
\usepackage{algpseudocode}
\usepackage{dirtytalk}
\usepackage{graphicx}
\usepgfplotslibrary{groupplots}
\usepackage{textcomp}
\usepackage{lineno}
\usepackage{svg}
\usepackage{optidef}
\usepackage{enumitem}
\usepackage{bbm}
\usepackage{setspace}
\usepackage{epsfig}
\usepackage{stfloats} 
\usepackage{array, multirow}
\usepackage{pgfplots}
\usepackage[subrefformat=parens,labelformat=parens,caption=false]{subfig}
\usepackage{stfloats}
\usepackage[lined,ruled,norelsize]{algorithm2e}
\makeatletter
\newcommand{\removelatexerror}{\let\@latex@error\@gobble}
\makeatother
\usepackage{titlesec}
\titlespacing{\subsection}{0.0pt}{\parskip}{-\parskip}
\usepackage{xcolor}
\def\BibTeX{{\rm B\kern-.05em{\sc i\kern-.025em b}\kern-.08em
    T\kern-.1667em\lower.7ex\hbox{E}\kern-.125emX}}
\begin{document}
\newcommand{\roa}[1]{\textcolor{blue}{[Ramoni: #1]}}
\title{Unsupervised Graph-based Learning Method for Sub-band Allocation in 6G Subnetworks\\
\thanks{The work by Daniel Abode was supported by the Horizon 2020 research and innovation programme under the Marie Skłodowska-Curie grant agreement No. 956670. The work by Gilberto Berardinelli, Renato Abreu, Thomas Jacobsen, and Ramoni Adeogun was supported by the HORIZON-JU-SNS-2022-STREAM-B-01-03 6G-SHINE project (grant agreement No. 101095738). Ramoni Adeogun's work was also partly supported by HORIZON-JU-SNS-2022-STREAM-B-01-02 project - CENTRIC (grant agreement No. 101096379)}
}
\author{\IEEEauthorblockN{Daniel Abode$^{1,2}$, Ramoni Adeogun$^{1}$, Lou Salaün$^{3}$, Renato Abreu$^{2}$, Thomas Jacobsen$^{2}$, Gilberto Berardinelli$^{1}$} \IEEEauthorblockA{\textit{$^{1}$Department of Electronic Systems, Aalborg University, Denmark.}\\ \textit{$^{2}$Nokia, Aalborg, Denmark. }\\ \textit{$^{3}$ Nokia Bell Labs, Massy, France} \\ Email:$^{1,2}$\{danieloa, ra, gb\}@es.aau.dk, $^{2}$renato.abreu@nokia.com, $^{2}$thomas.jacobsen@nokia.com, \\$^{3}$lou.salaun@nokia-bell-labs.com}} 

\maketitle
\begin{abstract}
In this paper, we present an unsupervised approach for frequency sub-band allocation in wireless networks using a graph-based learning method. We consider a scenario of dense deployment of subnetworks in the factory environment. The limited number of sub-bands must be optimally allocated to coordinate inter-subnetwork interference. Traditional iterative solutions may not scale to the large scale and density of subnetwork deployment due to their execution overhead limitations. Hence, we consider a data-driven approach based on graph neural networks. 
We model the subnetwork deployment as an interference graph and propose an unsupervised learning approach to optimize the sub-band allocation using graph neural networks. This approach is inspired by the graph colouring heuristic and the Potts model. The numerical evaluation shows that the proposed method achieves close performance to the centralized greedy colouring sub-band allocation heuristic with lower computational time complexity. In addition, it incurs reduced signalling overhead compared to iterative optimization heuristics that require all the mutual interfering channel information. We further demonstrate that the method is robust to different network settings. 
\end{abstract}


\section{Introduction}
The densification of wireless networks is necessary to support the growing connectivity demand, a trend that will continue towards sixth generation (6G) \cite{Zhang2019x}. The authors in \cite{VH2020} envisioned 6G to be a \say{network of networks}, where autonomous short-range low-power subnetworks inside entities such as vehicles, robots, industrial modules etc. can offload local communication within the entities from the central network. However, subnetwork deployments are expected to be extremely dense and large in scale in most use cases. For instance, in a factory scenario, numerous production modules and robots carrying subnetworks may be in proximity. The high density and large scale lead to significant inter-subnetwork interference, presenting a major limitation to the achievable spectral efficiency \cite{Adeogun2020}. Consequently, efficient methods including frequency allocation are being studied to mitigate the interference limitation \cite{Adeogun2023,Li2023,Saeed2024}. 

Interference mitigation via frequency allocation typically involves dividing the available bandwidth into sub-bands, or frequency channels to be reused at different cells in the network. This essentially prevents adjacent cells from operating on the same sub-band or channel. Frequency reuse is crucial to balance the tradeoff between maximizing the cellular network's ability to accommodate more users and mitigating interference \cite{Chang2009}. However, the number of sub-bands is usually limited compared to the number of cells. Hence, optimally allocating the sub-bands results in a combinatorial optimization problem which is known to be NP-hard \cite{Chang2009}. In traditional cellular architecture, where cell deployment is coordinated and static, optimized fixed frequency reuse patterns are generally efficient \cite{Chang2009}. However, in subnetworks, wireless cells are installed inside entities including vehicles, humans and robots which are mobile. In this case, the fixed-frequency reuse pattern is insufficient. Consequently, various sub-optimal heuristics and data-driven solutions are being developed to cope with the dynamic nature of inter-subnetwork interference. In the literature, centralized schemes such as centralized graph colouring (CGC) \cite{Adeogun2020} and sequential iterative sub-band allocation (SISA) \cite{Li2023} have been shown to achieve a better gain in spectral efficiency than distributed schemes.

In \cite{Li2023}, the authors introduced SISA to address the issue of inter-subnetwork interference. They formulated an optimization problem to minimize the sum of the interference-to-signal ratio across all subnetworks. They developed a centralized solution that iteratively allocates sub-bands based on the channel information of the inter-subnetworks interference links and the intra-subnetwork desired links. However, the drawback of this method lies in the spectral cost associated with gathering the channel information of the mutual inter-subnetwork interference and the subsequent signalling of this information to the centralized agent responsible for executing the sub-band allocation algorithm. This cost increases quadratically with the number of subnetworks \cite{Saeed2024}, hence the method is unsuitable for large-scale deployment.

Alternatively, the sub-band allocation can be formulated as a graph colouring problem as in \cite{Caidan2016, Adeogun2020, Silard2022}, where Access Points (AP) or cells are represented as nodes in a graph and the edges model mutual interference. The graph is designed so that the chromatic number is either the same as the number of sub-bands or very close to it. The chromatic number of a graph is the number of colours required to colour the nodes of the graph such that no adjacent nodes have the same colour. The graph colouring problem defines the assignment of colours (sub-bands) to nodes (AP or cell) such that no nodes connected by an edge (i.e., adjacent nodes) are assigned the same colour \cite{Caidan2016}. Formulating sub-band allocation as a graph colouring problem is advantageous because wireless networks can be naturally represented as graphs. Additionally, graph colouring has been extensively studied, resulting in several heuristic solutions. The common solutions use a sub-optimal greedy colouring algorithm since exact solutions are typically exponential in complexity \cite{Kosowski2008}. One significant benefit of modelling the sub-band allocation as graph colouring is the decrease in signalling cost. This is because the construction of the input graph only requires information about a small subset of the strongest interfering neighbours of each subnetwork, rather than the full channel information of both the desired and interfering links required in SISA. 

In this paper, we propose a method inspired by the graph colouring heuristic. Our approach aims to reduce signalling costs while employing a data-driven approach for reduced runtime complexity. Recently, there has been a growing interest in using a data-driven approach to solve combinatorial optimization problems such as graph colouring \cite{Zhe2007,MS}. This is driven by the goal of reducing runtime costs, particularly for large-scale scenarios. We also consider an unsupervised data-driven approach to reduce training costs. Supervised learning approaches such as in \cite{Adeogun2021} require the generation of large datasets including ground truths for training, which is complex to generate in practice. However, notable advances have been made in unsupervised learning approaches that do not require the complex procedure of generating ground truth using heuristics. Unsupervised methods have been shown to effectively solve combinatorial optimization problems using graph neural networks (GNN) \cite{MS}. This paper presents a novel unsupervised graph-based learning method for sub-band allocation. The approach incorporates a loss function based on the Potts model which penalizes the allocation of the same sub-band to adjacent nodes. This loss function does not depend on channel gain information. We employ a Gated Graph Neural Network (GGNN) as the graph-based learning model. It has been demonstrated to efficiently learn the graph structure for making inferences in sequential and combinatorial problems \cite{Li}. The GGNN is permutation-equivariant, scalable to changes in the size of the wireless network, and robust to different channel measurements. In addition, it can be executed in a centralized or decentralized manner.  


Our major contributions are as follows;
\begin{itemize}
    \item We represent the subnetworks deployment as an interference graph based on information on the strongest interfering neighbours of each subnetwork.
    \item We model the sub-band allocation as a node classification task and propose an unsupervised graph-based learning approach inspired by the Potts model. 
    \item We conduct extensive simulations to evaluate the performance and complexity of the approach compared to heuristic benchmarks. Furthermore, we evaluate the generalizability of the approach to different numbers of subnetworks, deployment density and channel models.
\end{itemize}
The rest of the paper is structured as follows. The next section presents the subnetwork system model and the problem formulation for sub-band allocation. In Section III, we describe the proposed GGNN algorithm followed by the performance evaluation and the simulation assumption in Section IV. Finally, we give some concluding remarks and future directions in Section V.

\begin{figure}
    \centering
    \includegraphics[scale=0.4]{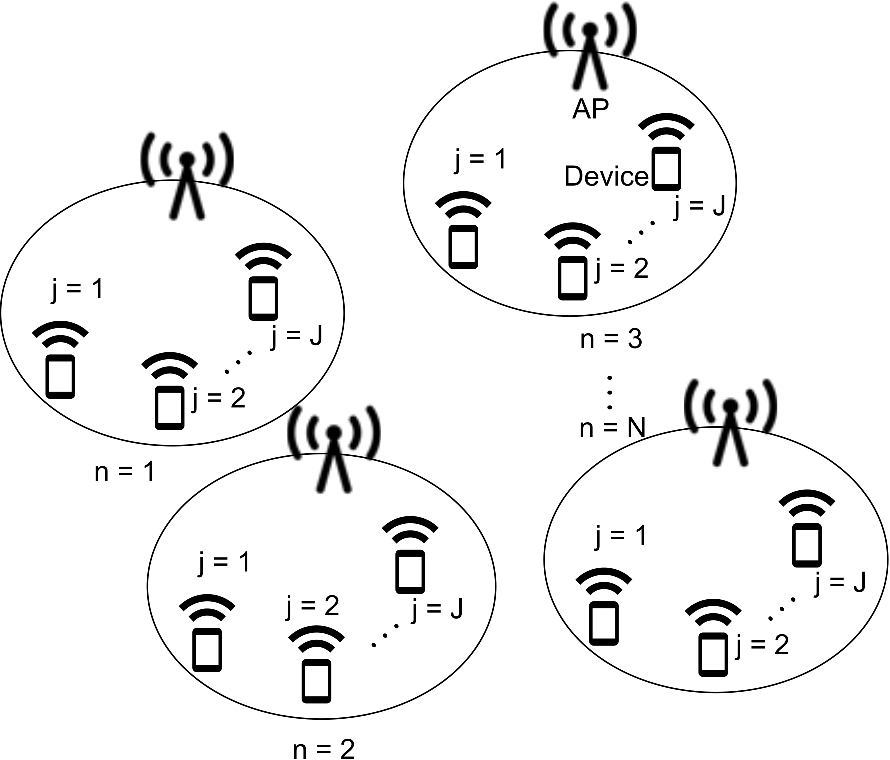}
    \caption{Deployment of N subnetworks with J devices}
    \label{fig:subnetwork}
\end{figure}

\section{System Model and Problem Formulation}
\subsection{Subnetworks System Model}
We consider a network of $\mathcal{N} = \{1,2,3,\cdots,N\}$ subnetworks which are densely and randomly deployed in an area as shown in Fig. \ref{fig:subnetwork} \cite{Adeogun2020}. Each subnetwork consists of an AP that coordinates the communication of its $\mathcal{J} = \{1,2,\cdots,J\}$ connected devices. The subnetwork and devices are indexed with $n \in \mathcal{N}$ and $j \in \mathcal{J}$ respectively. We assume that the subnetworks operate over a synchronized time-frequency resource grid with the available bandwidth divided into $K$ orthogonal sub-bands, where $K << N$. Hence, the $K$ sub-bands are expected to be reused by multiple subnetworks, generating mutual interference. The sub-bands allocated to a subnetwork are further partitioned into orthogonal time-frequency slots so that each device in a subnetwork is allocated a dedicated time-frequency slot to avoid intra-subnetwork interference. We assume that each subnetwork can be allocated only one sub-band $k \in \{1,2, \cdots, K\}$, which is identified by a one-hot encoded vector $\mathbf{\theta_n} \in \mathbb{B}^K$. Hence, for the network of $N$ subnetworks, we can define a sub-band selection matrix $\mathbf{\Theta} \in \mathbb{B}^{K \times N}$, such that $\mathbf{\theta_n} = \mathbf{\Theta}[:,n]$. The signal-to-interference noise ratio (SINR) of the uplink transmission between the $j$th device and the AP in subnetwork $n$ occupying a channel slot in sub-band $\mathbf{\theta_n}$, with a channel gain, $\gamma_{j,n}$ and fixed transmit power, $p_t$ can be written as

\begin{equation}
    \Gamma_{j,n} = \frac{p_{t}\mid \gamma_{j,n} \mid^2}{\sum\limits_{\substack{m=1 \\ m \neq n}}^N \mathbbm{1}{(\theta_n, \theta_m)} \ p_{t} \mid\gamma_{j',m,n}\mid^2  + \sigma^2},
    \label{eqn:SINR}
\end{equation}
\begin{equation}
    \mathbbm{1}{(\theta_n, \theta_m)} = 
    \begin{cases}
        1, & \text{if} \ \theta_n = \theta_m \\
        0, & \text{Otherwise},
    \end{cases}
\end{equation}

where $j'\in \mathcal{J}$ identifies the interfering device in subnetwork $m \in \mathcal{N}, m \neq n$ operating over the same time-frequency slot as device $j$ in subnetwork $n$ with the corresponding interfering channel gain, $\gamma_{j',m,n}$. $\sigma^2$ denotes the thermal noise power.

\vspace{-2.2pt}
\subsection{Interference Graph Model of Subnetworks}
The subnetwork deployment can be represented as a graph $G(\mathcal{V},\mathcal{E})$, where the set of nodes $\mathcal{V} = \{1, 2, \cdots, N\}$ represent the subnetworks and the set of edges $\mathcal{E} = \{(n,m) : n,m \in \mathcal{V}\}$ represents potential inter-subnetwork interference. An edge exists if subnetwork $n$ and subnetwork $m$ are considered neighbouring subnetworks, i.e. $m \in \mathbb{N}(n)$ which can be based on different rules, where $\mathbb{N}(n)$ denotes a set of the neighbours of $n$. Foremost, the resulting interference graph must have a chromatic number of $K$. That is, it should be possible that all the nodes in the subnetwork interference graph can be allocated orthogonal sub-bands such that no adjacent nodes are allocated the same sub-band given a maximum of $K$ sub-bands. One possible approximation to consider when building the subnetwork interference graph described in \cite{Adeogun2020} is to connect each subnetwork to $K-1$ strongest interfering neighbours. In this way, we consider a set $\mathcal{M}_{I_n}$ which includes all the $K-1$ strongest interfering subnetworks to subnetwork $n$. $G(\mathcal{V},\mathcal{E})$ is built, such that;
    \begin{equation}
        \{\forall(n,m) \mid \mathcal{E}_{n,m}  = 1 \text{ if } m \in \mathcal{M}_{I_n}, \text{ else } \mathcal{E}_{n,m}  = 0\},
    \end{equation}

\subsection{Sub-band Allocation as a Node Classification Task}
Given the subnetwork deployment interference graph $G(\mathcal{V},\mathcal{E})$, the node $n \in \mathcal{V}$ is labelled by the sub-band selection $\theta_n$. The optimization problem is to select $\theta_n, \theta_m$ such that $\mathbbm{1}{(\theta_n, \theta_m)} = 0$ if $m \in \mathbb{N}(n) \ \forall n$ which would intuitively minimize mutual inter-subnetwork interference. This optimization problem is similar to graph colouring \cite{Kosowski2008}. According to \cite{Zhe2007}, the Potts model on a graph $G(\mathcal{V},\mathcal{E})$ defines the Hamiltonian of the interaction between adjacent nodes $n,m$ with spin variables $\eta_n, \eta_m \in \{1,2,\cdots,K\}$ as
\begin{equation}
    \mathcal{H}(\eta) = \sum\limits_{(n,m) \in \mathcal{E}} \delta(\eta_n, \eta_m),
    \label{eqn:H}
\end{equation}
where, $\delta(\eta_n, \eta_m) = 0$ if $\eta_n \neq \eta_m$, which implies that the energy contribution of adjacent spins with different spin variables is zero, and positive otherwise. Essentially, if $G(\mathcal{V},\mathcal{E})$ is K-colourable, to achieve a ground state energy of zero, the model penalizes adjacent spins that have the same spin variables. In relation, we can consider the sub-band allocation $\theta_n$ as a one-hot code for a spin variable $\eta_n$ given the subnetwork interference graph. According to \cite{MS}, the Hamiltonian can be reformulated in terms of the one-hot vector $\theta_n$ to derive the loss function \cite{MS}
\begin{equation}
    \min_{\mathbf{\Theta}} \Psi(\theta_n) = \sum\limits_{(n,m) \in \mathcal{E}} \theta_n^T \cdot \theta_m .
    \label{eqn:loss1}
\end{equation}
This loss function would have a minimum value of zero if all the adjacent subnetworks are allocated orthogonal sub-bands, i.e. if $\theta_n \neq \theta_m \ \forall (n,m) \in \mathcal{E}$ corresponding to equation (\ref{eqn:H}). 
For the unsupervised procedure that minimizes (\ref{eqn:loss1}), we redefine the problem as a multi-class node classification, where a class is a sub-band. To enable a differentiable procedure, we replace the one-hot vector with a normalized smooth approximation of its class,  $\theta_n \mapsto \hat{\theta}_n \in [0,1]^K$. Hence,
\begin{equation}
    \min_{\mathbf{\hat{\Theta}}} \Psi(\hat{\theta}_n) = \sum\limits_{(n,m) \in \mathcal{E}} \hat{\theta}_n^T \cdot \hat{\theta}_m
    \label{eqn:loss2}
\end{equation}

The next section proposes the graph-driven model to learn $\hat{\theta}$ and describes the training procedure.

\section{Sub-band Allocation using Graph Neural Network}
\begin{figure}
    \centering
    \includegraphics[scale=0.43]{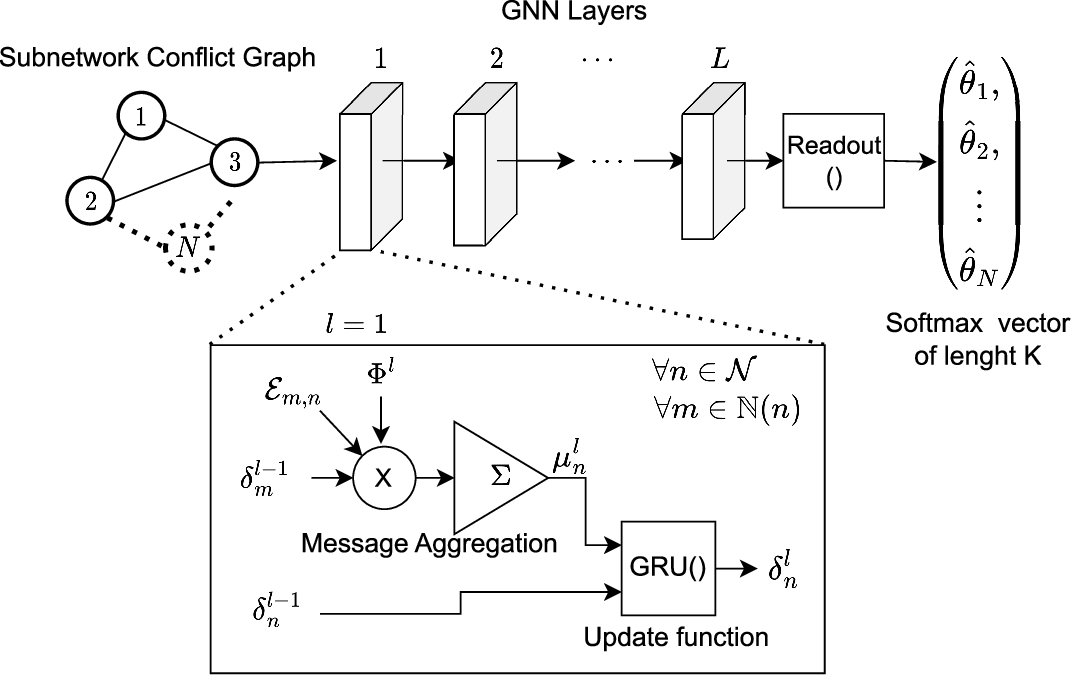}
    \caption{Graph Neural Network Design}
    \label{fig:GNN}
\end{figure}

GNN is a family of neural network algorithms capable of learning from graph signals. As shown in Fig. \ref{fig:GNN}, the GNN architecture consists of $L$ layers. The propagation model in a GNN layer can be described using two functions, the message aggregation function and the update function. The message aggregation function is a permutation equivariant function that aggregates a pairwise exchange of embeddings between adjacent nodes. The update function generates a new embedding for the node from the aggregated messages. We adopt the GGNN model \cite{Li} for the message aggregation and update functions. The GGNN architecture utilizes a gated recurrent layer as the update function.

\subsection{Gated Graph Neural Network Architecture}
Given the interference graph of the subnetwork deployment $G(\mathcal{V},\mathcal{E})$, the aggregated message for each node $n \in \mathcal{V}$ at layer $l \in \{1,2,\cdots,L\}$ as illustrated in Fig. \ref{fig:GNN} is given as;
\begin{equation}
    \mu_n^{l} = \sum\limits_{m\in\mathbb{N}(n)}\mathcal{E}_{m,n} \cdot \Phi^{l} \cdot \delta_m^{l-1},
    \label{eqn:agg}
\end{equation}
where $\Phi^{l}$ is the trainable feature transformation weight matrix of the message aggregation function, and $\delta_m^{l-1}$ denotes the previous node embedding of the neighbour. The update function is implemented using a gated recurrent unit (GRU), where the input is the aggregated message and the hidden state is the previous node embedding. So, the new embedding of node $n$ at layer $l$ is given as
\begin{equation}
    \delta_n^{l} = \text{GRU}(\mu_n^{l},\delta_n^{l-1}).
    \label{eqn:update}
\end{equation}
The GRU function consists of the reset gate $r^{l}$, update gate $u^{l}$ and the new gates $o^l$ which are define as
\begin{gather}
    \begin{gathered}
            r^l = \sigma(A_r^l\mu_n^{l} + a_r^l + B_r^l \delta_n^{l-1} + b_r^l), \\
    u^l = \sigma(A_u^l\mu_n^{l} + a_u^l + B_u^l \delta_n^{l-1} + b_u^l), \\
    o^l = \tau(A_o^l\mu_n^{l} + a_o^l + r^l \otimes (B_o^l\delta_n^{l-1} + b_o^l)),
    \end{gathered}
\end{gather}  
where $\sigma(\cdot)$ is a sigmoid activation function, $\tau(\cdot)$ is a hyperbolic tangent activation function. $A_r^l$, $B_r^l$, $A_u^l$, $B_u^l$, $A_o^l$, $B_o^l$ are trainable weights of the reset gate, update gate and new gate respectively, $a_r^l$, $b_r^l$, $a_u^l$, $b_u^l$, $a_o^l$, $b_o^l$ are their corresponding biases. The update function in (\ref{eqn:update}) is then given as
\begin{equation}
    \delta_n^{l} = (1 - u^l) \otimes o^l + u^l \otimes \delta_n^{l-1},
\end{equation}
where $\otimes$ denotes Hadamard product.

The trainable message aggregation weight allows the GGNN to learn the representation of the input graph structure while the update GRU allows the GGNN to learn the relationship between different intermediate internal embeddings in the $L$ layers \cite{Li}.

Finally, a readout function compresses the final node embedding after $L$ GGNN layers into a normalized soft vector of size $K$, $\hat{\theta_n}$ as in 
\begin{equation}
    \hat{\theta}_n = \text{Softmax}(W\delta_n^L + b),
    \label{eqn:theta}
\end{equation}
where $W$ and $b$ are the weights and biases of the readout function.

\subsection{Training and Execution Procedure}
We employed an unsupervised training algorithm which does not depend on any ground truth. The training graphs are generated using the interference graph model of the subnetworks. A mini-batch of graphs is propagated through the GGNN layers which execute (\ref{eqn:agg}), (\ref{eqn:update}), and (\ref{eqn:theta}). 
The output $\hat{\theta}_n \   \forall n$ is passed to the loss function (\ref{eqn:loss2}). By using mini-batch gradient descent, the $\hat{\theta}_n$ at training iteration $t$, $\hat{\theta}_n^t$ is updated as in

\begin{equation}
    \hat{\theta}_n^t = \hat{\theta}_n^{t-1} - \vartheta \mathbb{E}_\mathcal{B}\nabla_{\hat{\theta}_n}\Psi(\hat{\theta}_n^{t-1})
\end{equation}

The training is terminated when $\mathbb{E}_\mathcal{B}(\Psi(\hat{\theta}_n^{t}) - \Psi(\hat{\theta}_n^{t-1})) < \epsilon$, where $\epsilon$ is the error tolerance, $\mathbb{E}_\mathcal{B}$ is the expectation over the batch of graph, and $\vartheta$ is the learning rate. 

Since the input graphs have no attributes, all nodes are treated equally, hence the prediction depends on the structure of the graph which is learned during the message-passing procedure. The predicted sub-band for subnetwork $n$ is given as $\text{argmax}(\hat{\theta}_n)$.

The trained model can be executed in a centralized or decentralized manner. For decentralized execution, each subnetwork obtains a copy of the trained GGNN model and executes layer $l$ to obtain embedding $\delta_n^{l}$ based on the message $\delta_m^{l-1}$ received from neighbouring subnetworks. Hence, such implementation would require $L$ rounds of such message passing and the size of each message depends on the size of the embedding. This however would incur considerable signalling overhead and require synchronization between the subnetworks in executing each GGNN layer. On the other hand, centralized execution could be preferred if the subnetworks are within the coverage of a central network. In this case, the central network controller obtains the $K-1$ neighbour identifiers for all subnetworks, builds the interference graph, executes the trained GGNN model, and signals the sub-band selection decision to the subnetworks.

\section{Results and Discussion}
In this section, we discuss the simulation assumption for the subnetwork deployment, the GGNN model selection and training parameters shown in table \ref{tab:sim}, including brief descriptions of the benchmark algorithms. We compare the performance of the various benchmarks and our proposed approach in terms of the network spectral efficiency, execution complexity and generalizability. The spectral efficiency (SE) (bits/s/Hz) is approximated using Shannon capacity as $\text{SE}_{j,n} = \log_2 (1+ \Gamma_{j,n})$.

\subsection{Simulation Assumptions}
\subsubsection{Subnetwork Deployment}
To evaluate the proposed method, we randomly deploy $N=50$ subnetworks in a factory floor of size $40m \times 25m$ resulting in a density of $50000 \ \text{subnetworks/km}^2$. We consider $K=5$ sub-bands. To model the large-scale fading, we used the third generation partnership program (3GPP) technical report (TR) 38.901 Indoor Factory (InF) channel model \cite{3GPP} for the Dense-clutter Low-antenna (InF-DL) scenario and the associated model for the probability of non-line of sight (NLOS) and line of sight (LOS). The InF-DL scenario is appropriate since the subnetwork's AP and the devices are clutter-embedded. We consider full buffer uplink transmission. The transmission link path-loss, $\rho$ is represented by the alpha-beta-gamma model \cite{3GPP}. The shadow fading $s$ is modelled using the spatially correlated shadowing model used in \cite{Adeogun2023}. The small-scale fading is Rayleigh distributed and complex-valued, denoted as $h \sim \mathcal{CN}(0,1) $. Finally, the corresponding channel gain is then calculated as $\gamma = h \times \sqrt{10^{(\rho+s)/10}}$. 

\subsubsection{GGNN Model and Training Settings}
The GGNN model is implemented with PyTorch Geometric and comprises $L = 10$ layers, with each layer having an output node embedding of size 64. The model is trained with $50000$ graphs with a batch size of $64$ graphs for $500$ epochs, using the Adaptive Moment Estimation (ADAM) optimizer with an initial learning rate of $10^{-3}$. The parameters of the training settings and the GGNN model in Table \ref{tab:sim} were selected after conducting multiple experimental trials to determine the optimal configurations.

\begin{table}[]
\caption{Simulation Assumption}
\label{tab:sim}
\scalebox{0.58}{
\begin{tabular}{|l|l|l|l|}
\hline
\multicolumn{1}{|c|}{Parameter}  & \multicolumn{1}{c|}{Value} & \multicolumn{1}{c|}{Parameter}                        & \multicolumn{1}{c|}{Value} \\ \hline
Factory area                     & 40m x 25m      & Number of subnetworks, $N$                                 & 50                         \\ \hline
Subnetwork radius                & 2.5m                 & \multicolumn{1}{c|}{Number of devices per subnetwork} & 1                          \\ \hline
Minimum distance between APs     & 2.5m                 & Device to AP minimum distance                         & 1m                   \\ \hline
InF-DL clutter density, clutter size & 0.6, 2                     & Correlation distance                                  & 10m                        \\ \hline
Shadowing std (LOS, NLOS) & 4dB, 7.2dB                     & Path loss exponent (LOS, NLOS)                                 & 2.15, 3.57                        \\ \hline
Transmit power                   & 0 dBm                      & Number of sub-bands, $K$                                   & 5                          \\ \hline
Total bandwidth                  & 20 MHz                     & Center frequency                                      & 28 GHz                     \\ \hline
Noise figure                     & 10 dB                & Number of GGNN layers                                 & 10                         \\ \hline
Size of embedding                & 64                         & Training epochs                                       & 500                        \\ \hline
Training data size               & 50000                      & Batch size                                            & 64                         \\ \hline
Initial learning rate            & $10^{-3}$     & Optimizer                                             & ADAM                       \\ \hline
\end{tabular}}
\end{table}

\subsection{Benchmarks}
To evaluate the performance of our proposed scheme in improving the network performance, we compare it with the following schemes;
\begin{enumerate}
    \item Random Allocation (RA) - A distributed scheme where one sub-band is randomly selected from the available $K$ options for each subnetwork.
    \item Sequential Iterative Sub-band Allocation (SISA) - The sub-band selection sequential algorithm for subnetworks as detailed in \cite{Li2023} is a centralized iterative algorithm that minimizes the sum of the weighted interference.
    \item Centralized Graph Coloring (CGC) - The approach described in \cite{Adeogun2021} applies a greedy graph colouring heuristic for sub-band allocation in subnetworks. We applied this method to the same interference graph used for evaluating our proposed graph-based learning technique.  
\end{enumerate}

\subsection{Network Performance Evaluation}
\begin{centering}
    \begin{figure}
        \centering
        \input{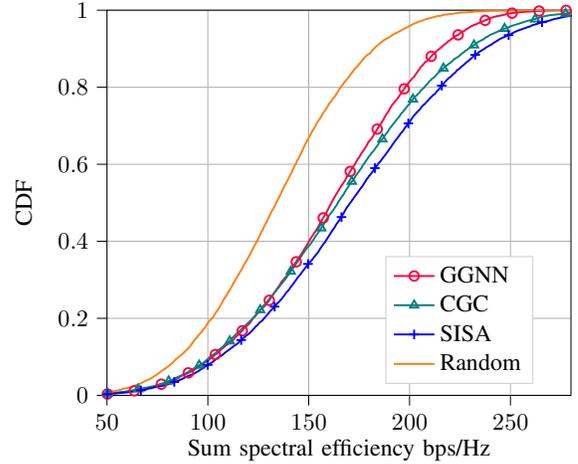}
        \caption{Cumulative distribution function (CDF) of the sum spectral efficiency for 10000 test snapshots}
        \label{fig:sumSE}
    \end{figure}
\end{centering}

\begin{centering}
    \begin{figure}
        \centering
        \input{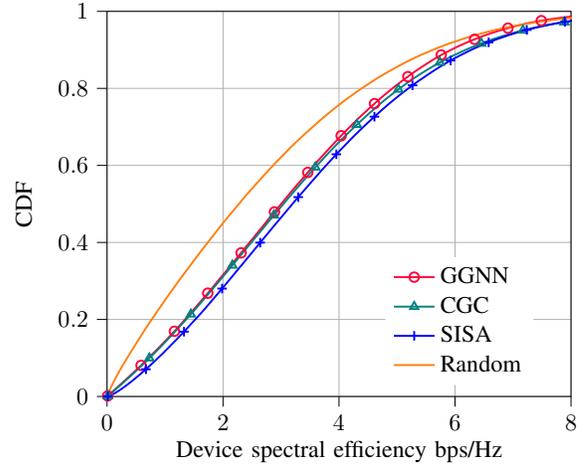}
        \caption{Cumulative distribution function (CDF) of the per-device spectral efficiency for 10000 test snapshots}
        \label{fig:indSE}
    \end{figure}
\end{centering}

Figures \ref{fig:sumSE} and \ref{fig:indSE} show the empirical cumulative distribution function (CDF) of the sum SE and per-device SE, respectively, for the proposed scheme and the different benchmarks tested with 10000 network realizations. Given the interference graph, our proposed method outperforms random sub-band allocation by $20\%$ in terms of achievable sum SE at the median. Below the median, GGNN achieves the same performance as CGC and lags behind SISA by $8\%$. However, note that SISA requires full channel gain information of all the mutual interfering links and desired links. Furthermore, GGNN can achieve a notable gain in per-device SE by up to $33\%$ compared to Random subband allocation at the median as shown in Fig. \ref{fig:indSE}. 

\begin{centering}
    \begin{figure}
        \centering
        \begin{tikzpicture}[scale=0.90]

\definecolor{darkgray176}{RGB}{176,176,176}
\definecolor{darkorange25512714}{RGB}{255,127,14}
\definecolor{forestgreen4416044}{RGB}{44,160,44}
\definecolor{steelblue31119180}{RGB}{31,119,180}
\definecolor{red}{RGB}{255,0,60}
\definecolor{blue}{RGB}{0,0,255}
\definecolor{lightgray204}{RGB}{255,255,255}

\begin{axis}[
legend cell align={left},
legend style={
  fill opacity=0.8,
  draw opacity=1,
  text opacity=1,
  at={(0.0,0.7)},
  anchor=south west,
  draw=lightgray204
},
tick align=outside,
tick pos=left,
ylabel = {Time(ms)},
x grid style={darkgray176},
xmajorgrids,
xmin=42.5, xmax=207.5,
xtick style={color=black},
y grid style={darkgray176},
xlabel = {Number of subnetworks},
ymajorgrids,
ymin=1.34199541807175, ymax=43.8010907769203,
ytick style={color=black}
]
\addplot [dashed, red, mark=o, mark size=3, mark options={solid}]
table {%
50 3.60772442817688
60 4.00972652435303
70 4.32450103759766
80 4.34334278106689
90 4.72414207458496
100 5.00572490692139
110 5.31423425674438
120 5.65466403961182
130 5.91401290893555
140 6.02741861343384
150 6.31173086166382
160 6.58063745498657
170 6.9404935836792
180 7.24755191802979
190 7.54800128936768
200 7.85946798324585
};
\addlegendentry{GGNN}

\addplot [dashed, teal, mark=triangle, mark size=3, mark options={solid}]
table {%
50 3.27195429801941
60 4.36027574539185
70 4.93752956390381
80 5.5459098815918
90 6.5501708984375
100 7.63566970825195
110 8.77511024475098
120 9.74986553192139
130 10.8464908599854
140 11.9308700561523
150 13.2322483062744
160 14.4715852737427
170 16.3326377868652
180 17.7110958099365
190 19.4075908660889
200 21.0409393310547
};\addlegendentry{CGC}

\addplot [dashed, blue, mark=+, mark size=3, mark options={solid}]
table {%
50 9.76164817810059
60 11.6957693099976
70 13.9707870483398
80 15.6113710403442
90 17.4726963043213
100 19.5271015167236
110 21.3734550476074
120 23.5450477600098
130 25.7252159118652
140 27.6036434173584
150 29.5222969055176
160 31.8150978088379
170 34.9156913757324
180 37.2278022766113
190 40.1598892211914
200 41.8711318969727
};\addlegendentry{SISA}

\end{axis}

\end{tikzpicture}
        \caption{Computational runtime for different number of subnetworks (ms) averaged over 10000 realizations}
        \label{fig:runtime}
    \end{figure}
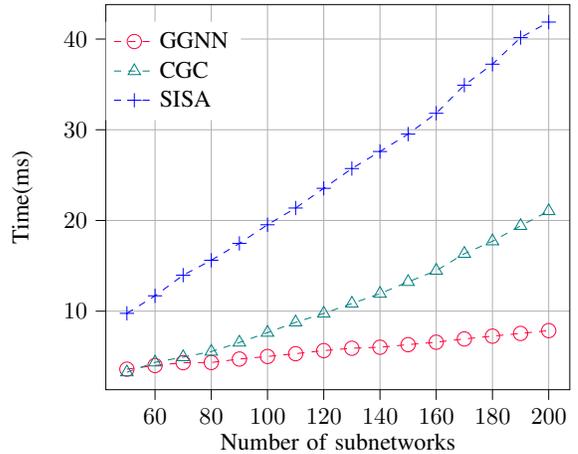
\end{centering}

\subsection{Complexity Analysis}
We analyzed the complexity of our proposed GGNN method and compared it with the benchmark algorithms, CGC and SISA in terms of the computational runtime and signalling requirement. Each algorithm is developed with Python frameworks and run on a Windows machine with an 11th Gen Intel(R) Core(TM) i7-11850H @ 2.50GHz processor and 32G memory. The result of the runtime analysis for different numbers of subnetworks, $N \in \{50,60,\cdots,200\}$ in Fig. \ref{fig:runtime} is averaged over $10000$ realizations. As shown in the figure, our GGNN method has a faster runtime, growing at a slower linear rate compared to the SISA and CGC. Hence, it would be more suitable for very dense networks with a large number of subnetworks or APs. While the runtime analysis is carried out on a CPU, it is expected that the runtime for the GGNN method would further decrease on a GPU.

For the signalling overhead considering centralized implementation, the GGNN requires less information and therefore incurs fewer signalling resources than SISA. For example, $N^2$ signalling messages are required to be signalled to the central resource management entity from $N$ subnetworks to execute the SISA algorithm; on the other hand, the GGNN method only requires $N(K-1)$ signalling messages, where $K<<N$ in a large network. This further justifies the suitability of the proposed method for a large-scale deployment of subnetworks.

\subsection{Generalizability}
We analyze the ability of the proposed graph-based learning method to generalize to different network settings and channel models different from the training system model assumption as shown in table \ref{table:Gen}. The default scenario is as presented in table \ref{tab:sim} and we consider two different scenarios. Scenario 1 is a network consisting of $80$ subnetworks deployed in a $50m \times 30m$ area, resulting in a $53300 \ \text{subnetworks}/km^2$ density, with the channel model based on 3GPP in-factory sparse clutter low antenna (InF-SL) model\cite{3GPP}. The NLOS path loss exponent, shadow fading standard deviation, clutter size and clutter density are $2.55$, $5.7$ dB, $10$ and $0.35$ respectively. Scenario 2 involves a less dense deployment of $20$ subnetworks in $25m\times25m$ area, i.e. $32000 \ \text{subnetworks}/km^2$ with path-loss modelled following the 3GPP model for inH-Office \cite{3GPP}. The NLOS path loss exponent, shadow fading standard deviation, and correlation distance are $3.83$, $8.03$ dB, and $6m$. We trained different GGNN models from training graphs for a given scenario and tested them on all three scenarios. As shown in table \ref{table:Gen}, we observe that the average SE from testing with 10000 snapshots remain relatively the same for a test scenario, regardless of the training scenario. This shows that the trained model can generalize to different numbers of subnetworks, density and channel models. The robustness to different settings is due to the fact that the GGNN model learns based on the graph structure, which depends on the graph construction rule and not the distribution of the channel model. On the other hand, a more detailed evaluation may be required to determine if the method can adapt to a much larger difference in density between the training and testing scenarios. It is important to note that if the number of sub-bands changes, a new GGNN model would be needed, since the size of $\hat{\theta}$ in \eqref{eqn:theta} corresponds to the number of available sub-bands.

\begin{table}[]
\caption{Generalizability of the proposed method to different network settings and channel model in terms of average rate ($\text{bps/Hz}$)}
\label{table:Gen}

\begin{tabular}{lcccc}
\multicolumn{2}{c}{}                                         & \multicolumn{3}{c}{Test}                                                                         \\ \cline{3-5} 
\multicolumn{1}{c}{}       & \multicolumn{1}{c|}{}           & \multicolumn{1}{c|}{Default} & \multicolumn{1}{c|}{Scenario 1} & \multicolumn{1}{c|}{Scenario 2} \\ \cline{2-5} 
\multicolumn{1}{l|}{}      & \multicolumn{1}{c|}{Default}    & \multicolumn{1}{c|}{2.9445}  & \multicolumn{1}{c|}{2.8167}     & \multicolumn{1}{c|}{6.1154}     \\ \cline{2-5} 
\multicolumn{1}{l|}{Train} & \multicolumn{1}{c|}{Scenario 1}  & \multicolumn{1}{c|}{2.9403}  & \multicolumn{1}{c|}{2.8124}     & \multicolumn{1}{c|}{6.1100}     \\ \cline{2-5} 
\multicolumn{1}{l|}{}      & \multicolumn{1}{c|}{Scenario 2} & \multicolumn{1}{c|}{2.9298}  & \multicolumn{1}{c|}{2.7965}     & \multicolumn{1}{c|}{6.1238}     \\ \cline{2-5} 
\end{tabular}
\end{table}

\section{Conclusion and Future Work}
This paper investigates an unsupervised graph-based learning approach to sub-band allocation for dense wireless subnetworks. The topology of the subnetwork is represented as a graph and the sub-band allocation is formulated as a node classification task parameterized by GGNN using a loss function inspired by the Potts model. We show that our approach offers comparable performance and requires lower runtime and signalling overhead than the centralized benchmark heuristics. We further show that the trained GGNN model is scalable, agnostic to the channel model and can be executed in a centralized or decentralized manner. Hence, the approach can be suitable for large-scale deployment of subnetworks. To better take advantage of the data-driven technique, our future work will consider more complex scenarios including traffic and mobility, where graph-based learning methods could outperform heuristics using the predictive ability of data-driven techniques.

\bibliographystyle{IEEEtran}
\bibliography{references}

\end{document}